\documentclass[sigconf]{acmart}
\AtBeginDocument{%
  }

\copyrightyear{2026}
\acmYear{2026}
\setcopyright{usgov}
\acmConference[HPDC '26]{The 35th International Symposium on High-Performance Parallel and Distributed Computing}{July 13--16, 2026}{Cleveland, OH, USA}
\acmBooktitle{The 35th International Symposium on High-Performance Parallel and Distributed Computing (HPDC '26), July 13--16, 2026, Cleveland, OH, USA}
\acmDOI{10.1145/3806645.3816239}
\acmISBN{979-8-4007-2640-8/2026/07}

\usepackage{enumitem}
\usepackage{balance}
\usepackage{tikz}
\usetikzlibrary{positioning,calc,arrows.meta}
\usetikzlibrary{shapes.geometric,shapes.misc}

\newcommand{\docfold}[1]{%
\fill[gray!15]
  ([xshift=1.5mm,yshift=1.5mm]#1.south east)
  --
  ([xshift=-1.5mm,yshift=1.5mm]#1.south east)
  --
  ([xshift=-1.5mm,yshift=-1.5mm]#1.south east)
  -- cycle;
\draw[gray!90]
  ([xshift=1.5mm,yshift=1.5mm]#1.south east)
  --
  ([xshift=-1.5mm,yshift=1.5mm]#1.south east)
  --
  ([xshift=-1.5mm,yshift=-1.5mm]#1.south east)
    --
  ([xshift=1.5mm,yshift=1.5mm]#1.south east);
}

\begin{document}

\title{Towards Transparent Checkpointing with AI-driven Code Generation}

\author{Hai Duc Nguyen}
\email{hai.nguyen@anl.gov}
\orcid{0000-0003-4177-0493}
\affiliation{%
  \institution{Argonne National Laboratory}
  \city{Lemont}
  \state{IL}
  \country{USA}
}

\author{Tekin Bicer}
\email{tbicer@anl.gov}
\orcid{0000-0002-8428-5159}
\affiliation{%
  \institution{Argonne National Laboratory}
  \city{Lemont}
  \state{IL}
  \country{USA}
}

\author{Kyle Chard}
\email{chard@uchicago.edu}
\orcid{0000-0002-7370-4805}
\affiliation{%
  \institution{University of Chicago}
  \city{Chicago}
  \state{IL}
  \country{USA}
}

\author{Ian Foster}
\email{foster@uchicago.edu}
\orcid{0000-0003-2129-5269}
\affiliation{%
  \institution{UChicago and Argonne}
  \city{Chicago}
  \state{IL}
  \country{USA}
}

\author{Bogdan Nicolae}
\email{bnicolae@anl.gov}
\orcid{0000-0002-0661-7509}
\affiliation{%
  \institution{Argonne National Laboratory}
  \city{Lemont}
  \state{IL}
  \country{USA}
}

\renewcommand{\shortauthors}{Nguyen et al.}

\begin{abstract}
  Adding reliable checkpoint/restart support to an MPI scientific application is a time-consuming expert effort that requires deep knowledge of both the application and resilience. We ask whether a frontier large language model can perform this work end-to-end without human intervention. We assemble a benchmark suite of MPI applications spanning diverse domains and computation patterns, and drive an iterative code-generation loop for each application using Anthropic's Claude Opus 4.7 invoked through the OpenCode CLI. Across six scientific applications, the LLM generates working checkpoint/restart code in 50~minutes on average while consuming 3.4~M tokens per application. The generated code adds negligible overhead during normal failure-free execution on five of six applications and recovers from injected process failures with efficiency comparable to human-engineered checkpoint/restart implementations. These results suggest that automated end-to-end LLM-driven resilience engineering is technically viable today for a meaningful fraction of HPC applications.
\end{abstract}



\maketitle

\thispagestyle{plain}
\pagestyle{plain}

\section{Introduction}
\label{sec:intro}

Modern scientific computing combines high performance computing (HPC) and
artificial intelligence (AI) in complex workflows that span heterogeneous infrastructure, from supercomputing centers to scientific instruments. This shift efficiently leverages heterogeneous accelerators and elastic capacity, but also
reshapes the failure landscape: long-running MPI jobs that once executed on a single homogeneous batch allocation now span resources whose mean-time-between-failure (MTBF) varies significantly; fluctuates dynamically as nodes are preempted, evicted, or reassigned under multi-tenant pressure; and is largely unpredictable at job-submission time~\cite{cappello2014toward,Diaspora-Overview-eScience24}. Under such circumstances, checkpoint/restart, the dominant resilience mechanism in HPC, faces several important challenges:
(1) which data structures within those components are critical and must be protected, and (2) at what point in the code it is safe to checkpoint those data structures. These challenges directly impact the correctness, performance, and scalability of checkpoint/restart solutions used to enable resilient execution.

Manual coding of checkpoint/restart solutions to address these challenges can be tedious due to the need to combine performance modeling;
deep understanding of interactions under parallelism to maintain a consistent
global state at scale; fine-tuning; and debugging. 
Recent advances in large language models (LLMs) and tool-using coding agents~\cite{opencode} offer a potential alternative that is both cheaper to implement and potentially more efficient.
The performance of these systems on bounded software-engineering benchmarks now approaches that of professional developers~\cite{jimenez2024swebench}, and their agentic workflows can autonomously read source trees, plan multi-step transformations, and react to validator feedback. 
We ask: \emph{can a frontier LLM coding agent be driven, end-to-end and with no human in the loop, to add production-grade checkpoint/restart resilience to unmodified MPI scientific applications, and at what cost in correctness, performance, and scalability?}

\textbf{Key Insights and Contributions.}
This paper investigates the operational viability of \emph{LLM-driven resilience engineering} as a building block for flexible scientific infrastructures. The observation we exploit is that the cognitive work of adding checkpoint/restart (identifying the data structures that make up the critical state, the correct place in the code where to checkpoint it, guarantee global consistency at scale, rebuild other data structures on restart based on the critical state) involves structured reasoning that modern coding agents are well suited for. The trade-off space, however, is far from understood: every iteration of the agent costs millions of tokens and tens of minutes of wall time; conservative state-coverage decisions inflate checkpoint footprint and erode scalability; a misidentified place in the code where it is safe to checkpoint silently corrupts a restart. Our goal is two-fold: (1)~establish empirically that frontier LLM coding agents can produce correct (w.r.t.\ failure-free execution output), performant resilience code with no human intervention; and (2)~characterize the cost, correctness and performance trade-offs that determine when this approach is practical for per-deployment resilience synthesis on dynamically provisioned, heterogeneous infrastructures. We summarize our contributions as follows:

\begin{enumerate}[itemsep=0pt,leftmargin=12pt]
\item \textbf{End-to-end LLM-driven resilience pipeline:}
We design a closed-loop generate--validate--repair pipeline that drives a frontier coding agent to add VeloC-based checkpoint/restart to an unmodified MPI source tree with no human in the loop. The validator builds the generated code, runs it failure-free, injects a single-rank fault, and structurally checks that the application restarts correctly and reproduces the vanilla output (\S~\ref{sec:setup}).

\item \textbf{Reusable resilience benchmark suite:} We assemble six MPI applications covering diverse domains, code sizes, state structures, and checkpoint back-ends. Each is shipped with a resilience-stripped \emph{vanilla} source (given to the agent) and an upstream-checkpointed \emph{reference} (human-engineered baseline) (\S~\ref{sec:suite}). 

\item \textbf{Empirical characterization of the cost, correctness and performance/scalability trade-off:} We quantify all three axes across the suite. On \emph{cost}, the agent converges in a median of two iterations, averaging 3.4~M tokens and 50~minutes of loop wall time per application. On \emph{correctness}, it consistently identifies a small, sufficient critical-state set and globally safe checkpoint points on every application. On \emph{performance}, the generated code imposes statistically zero failure-free overhead on five of six apps and recovers from injected faults within $\pm$6\% of the upstream human-engineered reference (\S~\ref{sec:findings}).
\end{enumerate}

\section{Related Work}

Application-level checkpointing approaches such as VeloC~\cite{veloc}, FTI~\cite{fti}, and SCR~\cite{scr} require developers to manually declare critical state, place checkpoint calls at safe points, and wire restart logic into initialization paths. By saving only the state needed for recovery, they keep checkpoints small and reduce both runtime and storage overhead. These systems typically rely on \emph{multi-level resilience strategies}~\cite{scr,VELOC-MASCOTS21}, where checkpoints are staged to a fast node-local tier, such as RAM, for recovery from frequent soft failures, then asynchronously flushed to a slower shared tier, such as a parallel file system, to survive critical failures.
A rich body of work further optimizes this pipeline: asynchronous flush engines overlap checkpoint I/O with application compute~\cite{DataStatesLLM-HPDC24,GPUPrefetch-HPDC23}; write aggregation coalesces per-rank writes into stripe-aligned bulk transfers~\cite{VELOC-FGCS24}; differential and incremental checkpointing persist only changed bytes~\cite{AICkptHPDC13,GPUDedup-ICPP23}; and compression or lossy encoding reduces footprint at modest reconstruction cost~\cite{SZ-CLUSTER19,LineageComp-HIPC23}. Checkpoint cadence is also optimized: the classical Young~\cite{young1974first}--Daly~\cite{daly2006higher} formulas minimize expected wasted work from MTBF and checkpoint cost, while adaptive variants track fluctuating MTBF on dynamically provisioned, multi-tenant resources~\cite{MLCkpt20}. These mechanisms reduce resilience cost at scale, but still presuppose the manual instrumentation step we automate: someone must decide what to checkpoint and where.

By contrast, transparent checkpointing tools such as BLCR~\cite{blcr}, DMTCP~\cite{dmtcp}, and CRIU~\cite{criu} reduce developer effort by capturing process state automatically, but often include caches, transient buffers, and rebuildable state. This can produce oversized checkpoints, limit scalability under elasticity, and complicate portability across heterogeneous nodes. Compiler-assisted approaches~\cite{bronevetsky2003} narrow the snapshot through static analysis, but require per-language passes and can struggle with the templated, dynamically allocated, irregularly accessed data structures common in modern HPC and AI codes. Although transparent and compiler-assisted approaches can benefit from the same multi-level optimizations, they do not provide the combination we target: low-effort automation with application-level checkpoint efficiency.

To our knowledge, no prior work has attempted to study frontier LLM code generation to the point where it allows end-to-end automation (i.e., analyze the application code and generate application-level checkpointing that captures a minimal critical state), thus combining the advantages of transparent checkpointing with application-level checkpointing.

\section{Resilient Benchmark Suites}
\label{sec:suite}

To test the effectiveness of frontier LLM models with respect to automated resilience, we assembled a benchmark suite of unmodified MPI scientific applications drawn from a wide spectrum of HPC domains. Three selection criteria apply: (i)~each application must be open source and reproducible from its upstream repository; (ii)~each must ship with a native checkpoint/restart implementation we can use as the upstream-reference baseline; and (iii)~together the suite must cover diverse data-structure shapes (fixed-size arrays, variable-per-rank particle pools, adaptive meshes), per-step synchronization patterns, and checkpoint back-ends.

\begin{table}[t]
\centering
\caption{Benchmark applications analyzed in this paper.}
\label{tab:apps}
\small
\setlength{\tabcolsep}{4pt}
\begin{tabular}{@{}lllr@{}}
\toprule
App & Lang & Checkpoint mech. & LOC \\
\midrule
Athena++  & C++ & native        & 132K \\
CoMD      & C   & POSIX         & 5.6K \\
HPCG      & C++ & POSIX         & 6.2K \\
LAMMPS    & C++ & native        & 627K \\
OpenLB    & C++ & native        & 338K \\
SPARTA    & C++ & native        & 174K \\
\bottomrule
\end{tabular}
\end{table}

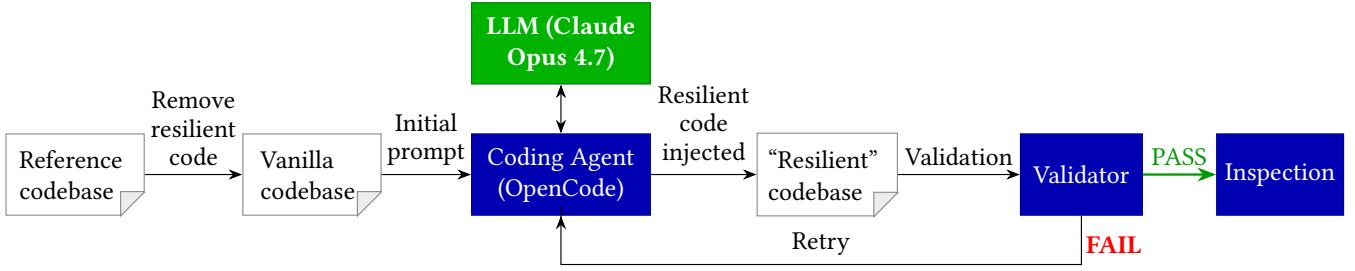
\begin{figure*}[hbtp]
    \centering
    \resizebox{\textwidth}{!}{%
\begin{tikzpicture}[
    font=\normalsize,
    >=Stealth,
    node distance=1.8cm and 1.8cm,
    doc/.style={
        draw=gray!90,
        fill=white,
        minimum width=1.7cm,
        minimum height=1cm,
        inner xsep=1pt,
        align=left,
        chamfered rectangle,
        chamfered rectangle corners={south east},
        chamfered rectangle xsep=2mm
    },
    agent/.style={
        draw=blue!60!black,
        fill=blue!70!black,
        text=white,
        minimum width=2.2cm,
        minimum height=1cm,
        align=center
    },
    validator/.style={
        draw=blue!60!black,
        fill=blue!70!black,
        text=white,
        minimum width=1.5cm,
        minimum height=1cm,
        align=center
    },
    llm/.style={
        draw=green!50!black,
        fill=green!70!black,
        text=white,
        minimum width=2.2cm,
        minimum height=1cm,
        align=center,
        font=\bfseries
    }
]

\node[doc] (ref) {\hspace*{-1mm}Reference\\\hspace*{-1mm}codebase};
\docfold{ref}
\node[doc, right=1.2cm of ref] (vanilla) {\hspace*{-1mm}Vanilla\\\hspace*{-1mm}codebase};
\docfold{vanilla}

\node[agent, right=1.1cm of vanilla] (agent) {Coding Agent\\(OpenCode)};
\node[llm, above=0.6cm of agent] (llm) {LLM (Claude\\Opus 4.7)};

\node[doc, right=1.3cm of agent] (res) {\hspace*{-1mm}``Resilient''\\\hspace*{-1mm}codebase};
\docfold{res}

\node[validator, right=1.5cm of res] (validator) {Validator};
\node[validator, right=0.9cm of validator] (inspect) {Inspection};

\draw[->] (ref) -- node[above,align=center] {Remove\\[-1pt]resilient\\[-1pt]code} (vanilla);
\draw[->] (vanilla) -- node[above,align=center] {Initial\\[-1pt]prompt} (agent);
\draw[<->] (llm) -- (agent);
\draw[->] (agent) -- node[above,align=center] {Resilient\\[-1pt]code
\\[-1pt]injected} (res);
\draw[->] (res) -- node[above] {Validation} (validator);
\draw[->,green!60!black,thick]
    (validator) -- node[above,text=green!60!black] {PASS} (inspect);

\draw
    ($(validator.south)+(0.4,-0.35)$)
    node[text=red,font=\bfseries] {FAIL};
\coordinate (retrylevel) at ($(validator.south)+(0,-0.6)$);
\draw[->]
    (validator.south) |- (retrylevel)
    -- node[midway,above] {Retry}
       (agent.south |- retrylevel)
    |- (agent.south);
\end{tikzpicture}
}
    \caption{End-to-end LLM-driven pipeline for adding checkpoint/restart resilience to an
             unmodified MPI application.}
    \label{fig:code-gen-process}
\end{figure*}

We identified six applications, listed in \autoref{tab:apps}, that meet these criteria. These applications 
span six computational domains and present distinct checkpointing challenges:
\begin{enumerate}[itemsep=0pt,leftmargin=*]
    \item \textbf{HPCG}~\cite{hpcg}: a conjugate-gradient sparse linear solver whose main loop runs a fixed number of independent CG solves and accumulates a residual after each one.
    \item \textbf{CoMD}~\cite{comd}: a classical molecular dynamics proxy in which atoms move under pairwise forces and may migrate between MPI subdomains each timestep.
    \item \textbf{OpenLB}~\cite{openlb}: a lattice-Boltzmann fluid solver whose central per-rank object is a large lattice with built-in buffer serialization.
    \item \textbf{SPARTA}~\cite{sparta}: a direct-simulation Monte Carlo code for rarefied gases in which particles drift between ranks every timestep, leaving each rank with a different particle count.
    \item \textbf{Athena++}~\cite{athenapp}: an astrophysical magnetohydrodynamics code with a mesh topology that evolves via adaptive mesh refinement.
    \item \textbf{LAMMPS}~\cite{lammps}: a production molecular dynamics code with extensive topological connectivity and diverse plugin state.
\end{enumerate}
Together they mix C and C++ codebases, fixed-size and variable-size per-rank state, and both POSIX and native checkpoint back-ends, exercising the principal checkpoint patterns encountered in production MPI software.

\section{Methodology}
\label{sec:setup}

\medskip
\noindent\textbf{Code Generation Pipeline.}
Figure~\ref{fig:code-gen-process} sketches the end-to-end pipeline to test whether
a frontier LLM can autonomously add checkpoint/restart resilience to an MPI scientific
code.  We strip the upstream source of any existing resilience logic to
produce a \emph{vanilla codebase}; this is the only artifact the LLM ever sees.  We then
drive an iterative code-generation loop, in which each iteration executes two phases in
sequence with no human in the loop:

\begin{enumerate}[itemsep=0pt,leftmargin=*]
    \item \textbf{Code Generation.} We give Anthropic's Claude Opus 4.7~\cite{claude}, invoked through the \texttt{opencode} coding-agent CLI~\cite{opencode}, the vanilla source tree plus a fixed instruction prompt that asks it to use VeloC~\cite{veloc} (a multi-level checkpoint/restart runtime designed for HPC) to protect the application against process failure. The agent issues \texttt{read}/\texttt{edit} tool calls on the source tree and writes a modified copy we call the \emph{resilient codebase}, which must preserve the vanilla code's functional behavior while adding fault tolerance. The prompt also asks the agent to log its reasoning, intended action, observed result, and next step at every iteration, giving us a per-step transcript of the transformation.

    \item \textbf{Validation.} An independent validator builds the resilient codebase and, if success, runs it under two scenarios. \emph{Failure-free execution} confirms that the modified code still produces output equivalent to the vanilla, ensuring the functional contract is intact. \emph{Failure-injected execution} kills one rank mid-run, re-launches the binary, and checks four conditions: (i)~at least one checkpoint file was written before the kill; (ii)~the harness delivered the failure (no spurious early exit); (iii)~the post-restart output matches the vanilla, confirming the checkpoint and recovery path reproduces the intended results; and (iv)~end-to-end wall time stays meaningfully below a full restart-from-scratch, evidence the checkpoint cadence is operationally useful.
\end{enumerate}

\noindent
If any check fails or the run times out, the loop iterates: the agent receives the original prompt, its previous resilient code, and the validator's structured feedback (build log, execution stdout/stderr, per-condition pass/fail flags) and is asked to produce a corrected version. Each application is allowed up to ten iterations; if the validator passes before that cap, the resulting code is promoted to the inspection phase analyzed in \S\ref{sec:findings}.

\medskip
\noindent\textbf{Metrics and Configurations.}
The inspection phase collects four metrics per application:
(i)~\emph{Code generation time}: cumulative wall time of the LLM/validator loop, including both the agent's thinking time and the validator's build/run time;
(ii)~\emph{Total tokens}: input plus output tokens consumed across all iterations, a direct proxy for API spend;
(iii)~\emph{End-to-end execution time}: wall time of one application run under each scenario, quantifying both instrumentation overhead and recovery cost; and
(iv)~\emph{Per-frame checkpoint footprint}: bytes written by a single checkpoint event, capturing the storage cost of the chosen state coverage.
All measurements were collected on a single development host with four MPI ranks per application. Per-app input arguments were tuned so one failure-free run takes 60--200~s, short enough to keep the iterative loop tractable yet long enough for failure injection to produce meaningful checkpoint, recovery, and output signals. For every (application, scenario) cell we run three independent trials and report the mean. All runs use the latest released version of VeloC~\cite{veloc}.

\section{Key Findings}
\label{sec:findings}

We trace what actually happens inside the iterative loop, distilled from the per-iteration \texttt{opencode\_stdout} transcripts that record the agent's reasoning and actions step by step.  Across all six applications the transcripts reveal the \emph{same four-phase procedure}, applied with few app-specific deviations:

\begin{enumerate}[itemsep=0pt,leftmargin=12pt]
    \item \textbf{Reconnaissance.}  The agent locates the application's main timestep loop, starting from the entry-point file (\texttt{main.cpp} or equivalent) and following function calls inward.  It also locates the VeloC installation on disk and reads the VeloC documentation to learn the API it will need.
    \item \textbf{Critical-state identification.}  The agent walks the application's central data structure field by field and labels each field as either \emph{rebuildable} (the input deck plus the application's own initialization code can recreate it) or \emph{evolving} (the value changes inside the timestep loop and therefore must be saved).
    \item \textbf{Implementation.} The agent wires VeloC into the application: it initializes the runtime at startup, declares memory regions of evolving fields to checkpoint, calls the checkpoint primitive at a chosen cadence inside the timestep loop, and adds a startup branch that detects an existing checkpoint, restores the saved state, and resumes the loop from the saved counter.
    \item \textbf{Correction.} When an implementation fails validation, the agent uses the validator’s failure report, as its primary feedback. Each subsequent iteration begins by analyzing this report, forming a single hypothesis about the likely source of the fault, and applying a targeted fix.
\end{enumerate}

\begin{figure}[t]
  \centering
  \includegraphics[width=\columnwidth]{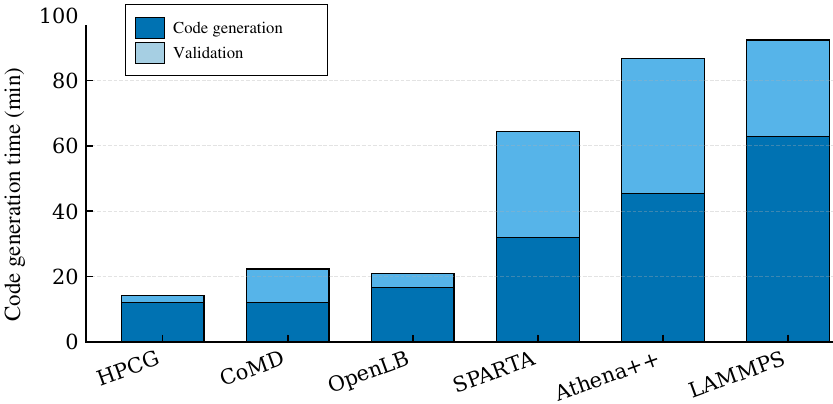}
  \caption{Time consumed by the LLM iterative loop to produce a working checkpoint/restart implementation, per app.
  }
  \label{fig:walltime}
\end{figure}

This procedure successfully transformed \textit{all} six vanilla codebases into resilient ones. We walk through each application below to show how the procedure plays out against the codebase's structure.

\paragraph{HPCG (2 iterations to PASS)}
HPCG is the suite's simplest case: each CG solve starts from the same zeroed initial vector and no state carries over between solves, so the only critical state is a counter plus the residual array. The agent rejected saving the sparse matrix, multigrid hierarchy, and per-solve scratch vectors as deterministically rebuildable from the input dimensions.

\paragraph{CoMD (2 iterations to PASS)}
CoMD is harder than HPCG because state evolves inside the timestep loop: atoms migrate between MPI subdomains every step, so the per-rank atom list (positions, momenta, forces, identifiers) plus a small link-cell bookkeeping array changes continuously. Everything else (simulation box, species table, potentials, link-cell geometry) is fixed at startup, and halo cells are repopulated on the first post-restart timestep by CoMD's own halo-exchange. The agent saved the atom arrays plus the loop counter and skipped halos.

\paragraph{OpenLB (1 iteration to PASS)}
OpenLB stores far more data per process than CoMD, and that data is tangled inside an internal object whose layout would be tedious and error-prone to list by hand. Conveniently, OpenLB already provides three helper methods designed for sending the object between processes (return size, pack into a buffer, unpack from a buffer); after an intensive search over the codebase, the agent found and reused them as a black-box serialization, so it never needed to know what the object contains.

\paragraph{SPARTA (2 iterations to PASS)}
SPARTA introduces a new wrinkle: per-rank particle counts vary because particles drift between ranks every timestep, ruling out fixed-size memory regions. The agent used VeloC's file-based API to write a small structured payload per rank (timestep counter, simulation-time scalars, particle count, particle array, RNG state) and skipped the static grid topology, species tables, and master RNG.

\paragraph{Athena++ (4 iterations to PASS)}
Athena++ raises the stakes again: its state lives in an evolving AMR mesh tree plus a long list of conditional physics arrays, so field-by-field enumeration would have been both labor-intensive and high-risk. The agent instead implemented a writer/reader pair dedicated to Athena++'s mesh format to dump and reconstruct the critical state from disk.

\paragraph{LAMMPS (7 iterations to PASS)}
LAMMPS has the most complex state in the suite. Its atoms move between ranks as in SPARTA, but it also tracks two things no other app does: bonds and angles that link atoms together (a graph the checkpoint must keep consistent after migration), and many small pieces of state owned by whichever optional physics modules the user enables (each keeps its own counters and random-number state).

\paragraph{Cross-cutting observations}
Two interesting patterns recur across all six apps. First, the agent always starts from a \emph{small} state-coverage set rather than a conservative \emph{everything} one: it picks the few fields that actually evolve, defends every rejection in writing, and only adds coverage if validator feedback forces it. This is the proximate reason for resulting checkpoints being small enough to write without measurable runtime overhead. 
Second, after identifying the critical state, the agents often do not proceed directly to implementation. Instead, they perform additional code reading to find and reuse existing utilities for handling critical-state save/load operations, such as serialization and file I/O. This helps them generate cleaner and lower-risk code.

\medskip
\noindent\textbf{Time to Construct Resilient Code.}
The iterative loop consumed 5.0~hours total across the six apps, averaging 50~minutes per app, with a $\sim$6.5$\times$ spread between the fastest (HPCG, 14~min) and the slowest (LAMMPS, 92~min). Figure~\ref{fig:walltime} decomposes per-app loop time into LLM code generation (dark blue, bottom) and validation (light blue, top): \emph{code generation dominates on every app}, taking a median of 61\% of the total (range 50--84\%).
The bar shapes track the difficulty ordering. HPCG (14~min) and CoMD (22~min) sit at the bottom because their critical state is small and fixed-size per rank, so the first iteration design was structurally correct and the second iteration only fixed a one-line operational glitch. SPARTA (64~min) sits in the middle. Athena++ (87~min) and LAMMPS (92~min) dominate the chart because their richer state surfaces forced multiple iterations to capture and recover correctly.

\begin{figure}[t]
  \centering
  \includegraphics[width=\columnwidth]{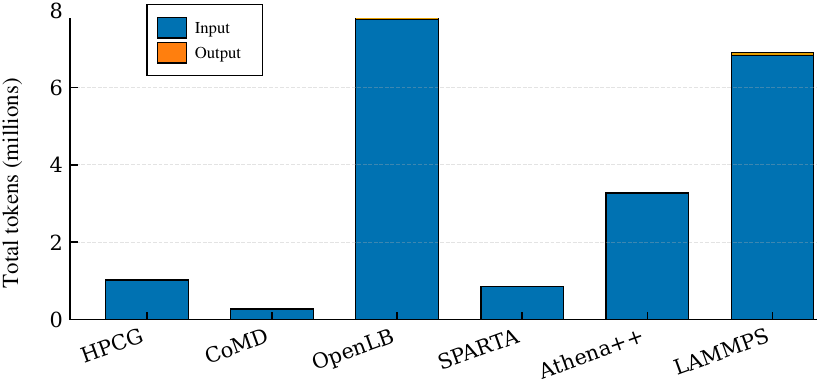}
  \caption{Total Claude Opus 4.7 tokens consumed by the iterative loop, per app (Output tokens are negligible).
  }
  \label{fig:tokens}
\end{figure}

\medskip
\noindent\textbf{Token Cost.}
Total tokens across the six apps sum to 20.2~M (average 3.4~M per app), corresponding to roughly \$50--400 in API spend at current frontier-model rates \cite{anthropic_pricing}. \autoref{fig:tokens} spans $\sim$28$\times$, from CoMD's 0.28~M up to OpenLB's 7.81~M.
Token cost correlates only loosely with code generation time. OpenLB tops the chart at 7.81~M despite completing in a single iteration, because the agent had to scan a substantial portion of OpenLB's 338K-line source tree to discover the built-in serialize methods for checkpoint serialization. CoMD sits at the bottom (0.28~M) because of its small codebase and critical state. LAMMPS at 6.91~M is large despite a small state struct because each iteration re-read the integrator and atom-management sources.

The stacked decomposition shows that input tokens account for roughly 99\% of every bar (19.98~M input vs.\ 0.19~M output across the suite). \emph{Input-context size is the dominant token cost driver}, so pre-summarizing or chunking large source trees would translate directly into spend reductions, especially on apps such as OpenLB where exploration substantially exceeded eventual edits.

\begin{figure}[t]
  \centering
  \includegraphics[width=\columnwidth]{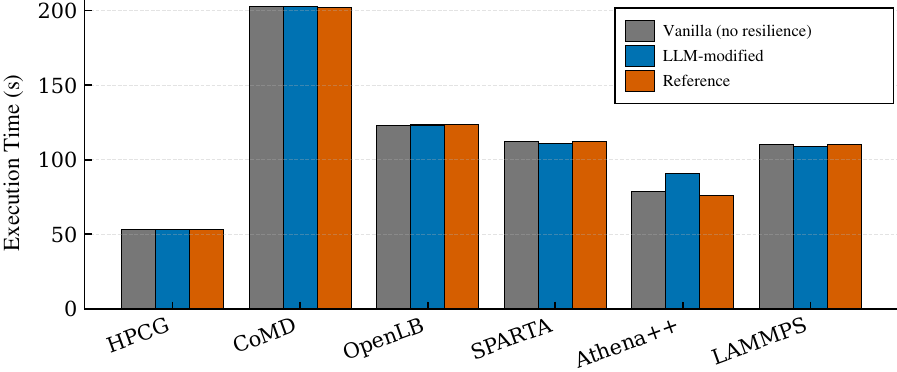}
  \caption{Per-app average execution time under the failure-free scenario.}
  \label{fig:ff}
\end{figure}

\begin{figure}[t]
  \centering
  \includegraphics[width=\columnwidth]{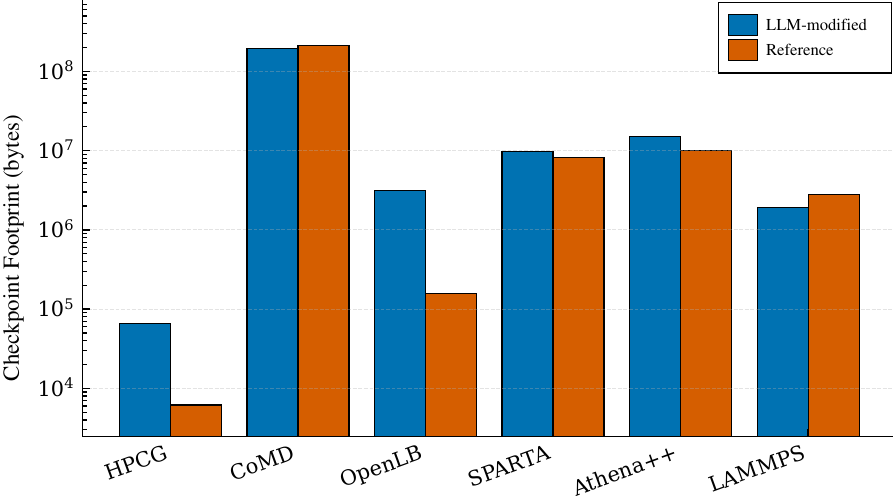}
  \caption{Payload size of a full recoverable checkpoint.}
  \label{fig:frame}
\end{figure}

\medskip
\noindent\textbf{Overhead: Runtime and Storage.}
\autoref{fig:ff} shows an important result: \textit{the LLM-generated checkpoint instrumentation imposes essentially zero overhead}. Five of six apps fall within $\pm$2\% of vanilla wall time (from -1.4\% for LAMMPS to +0.2\% for CoMD and OpenLB), inside measurement noise. The outlier is Athena++ at +15.2\%, where the agent chose an aggressive 8\,s cadence that trades failure-free performance for faster recovery. The LLM bars also track the upstream reference closely on every app, showing that the generated code matches a human-engineered native implementation.

\autoref{fig:frame} shows full-checkpoint sizes spanning four orders of magnitude, from HPCG's 66~KB (residual array plus a few scalars) up to CoMD's 194~MB (full per-rank atom array). LLM and reference agree on the order of magnitude for four of six apps (CoMD, SPARTA, Athena++, LAMMPS). On the two where they diverge (HPCG: LLM~66~KB vs.\ reference~6~KB; OpenLB: LLM~3~MB vs.\ reference~157~KB) the LLM is more conservative, persisting diagnostic accumulators the reference recomputes on restart.

\begin{figure}[t]
  \centering
  \includegraphics[width=\columnwidth]{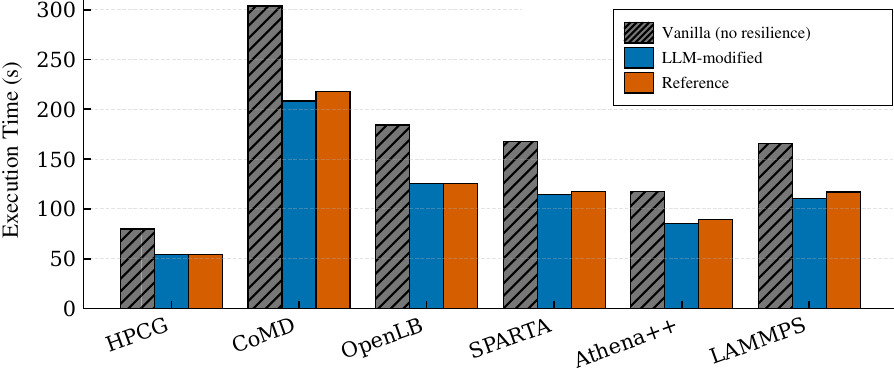}
  \caption{Per-app end-to-end execution time with a failure injected mid-run.
  }
  \label{fig:once}
\end{figure}

\medskip
\noindent\textbf{Resilience.}
\autoref{fig:once} shows per-app end-to-end runtime when the validator kills one MPI rank mid-run, then restarts the app. The LLM runtime sits close to the reference on every app, ranging from 6\% faster (Athena++) to 3\% slower (CoMD, SPARTA), and both are roughly one third faster than the vanilla baseline, which has no checkpoint protection and pays the full 50\% wall-clock penalty of restarting from scratch. This is operational evidence the LLM-generated code does real recovery work: detecting checkpoint files, restoring state, and resuming with overhead bounded by the checkpoint cadence rather than by the size of the original computation.

\section{Conclusion and Future Work}
\label{sec:conclusion}

This paper investigates whether frontier LLM coding agents can automate end-to-end checkpoint/restart integration for existing MPI scientific applications. Across six applications with diverse domains, code sizes, data structures, and checkpointing patterns, the agent generated working resilient implementations without human intervention after the initial prompt. The generated code achieved near-zero failure-free overhead on five of six applications, recovered from injected failures with performance close to upstream native implementations, and converged in under two hours per application. These results suggest that LLM-driven resilience engineering can substantially reduce the expert effort required to add production-grade fault tolerance to a meaningful subset of HPC applications.

Future work should improve the robustness, scalability, and usability of this approach. Specialized pipelines with resilience-specific skills, source-code summarization, checkpoint-state analysis, and targeted validation feedback could reduce cost and improve reliability. Broader evaluations are needed for repeated failures, realistic I/O environments, multi-node executions, and applications with more irregular or distributed state. Finally, generated checkpoint code should become easier to audit, maintain, and integrate into production workflows so that automated resilience generation can be used safely in practice.

\begin{acks}
This work was supported in part by U.S.\ Department of EnergyContract DE-AC02-06CH11357, and National Science Foundation grants CSSI-2411386 and CSSI-2514056.
\end{acks}

\balance
\newpage
\bibliographystyle{ACM-Reference-Format}
\bibliography{references}

\end{document}